\documentclass{PoS}

\usepackage{amsmath}
\usepackage{amsfonts,amsbsy,amsxtra}
\usepackage{amssymb,latexsym}
\usepackage{textcomp}
\usepackage{ifthen}

\newcommand{\figref}[1]{{Fig.~\ref{fig:#1}}}
\newcommand{\Figref}[1]{{Figure~\ref{fig:#1}}}
\newcommand{\Figrefs}[1]{{Figures~\ref{fig:#1}}}
\newcommand{\figrefshort}[1]{{\ref{fig:#1}}}

\newcommand{\others}{\textit{et al.}}

\newcommand{\antibar}[1]{%
  \overline{#1}%
}
\newcommand{\PPh}{\ensuremath{\gamma}}

\renewcommand{\Pr}{\ensuremath{\rho}}
\newcommand{\Prthree}{\ensuremath{\rho_3}}

\newcommand{\Paone}{\ensuremath{a_1}}
\newcommand{\Patwo}{\ensuremath{a_2}}

\newcommand{\Pfzero}{\ensuremath{f_0}}

\newcommand{\Pftwo}{\ensuremath{f_2}}

\newcommand{\Ppi}{\ensuremath{\pi}}
\newcommand{\Ppione}{\ensuremath{\pi_1}}
\newcommand{\Ppitwo}{\ensuremath{\pi_2}}

\newcommand{\Ppip}{\ensuremath{\pi^+}}
\newcommand{\Ppim}{\ensuremath{\pi^-}}

\newcommand{\PKm}{\ensuremath{K^-}}

\newcommand{\Ppbar}{\ensuremath{\antibar{p}}}

\newcommand{\PA}{\ensuremath{A}}
\newcommand{\PPb}{\ensuremath{\text{Pb}}}
\newcommand{\twopion}{\ensuremath{\Ppip\Ppim}}
\newcommand{\threepion}{\ensuremath{\Ppim\Ppip\Ppim}}

\newcommand{\abs}[1]{{|{#1}|}}

\newcommand{\measresult}[4]{%
  \ensuremath{#1%
    \ifthenelse{\equal{#2}{}}%
    {}%
    {\pm #2%
      \ifthenelse{\equal{#3}{}}%
      {}%
      {_\text{stat.}}%
    }%
    \ifthenelse{\equal{#3}{}}%
    {}%
    {\pm #3_\text{syst.}}\text{#4}%
  }%
}

\newcommand{\jpc}{\ensuremath{J^{PC}}}

\newcommand{\wavespec}[7]{\ensuremath{#1^{#2#3}}\!\!
  \ensuremath{#4^{#5}}\!\! \ensuremath{[#6] #7}} 

\newcommand{\gevc}{~\ensuremath{\text{GeV}\! / c}}
\newcommand{\gevcsq}{~\ensuremath{(\text{GeV}\! / c)^2}}
\newcommand{\mevcc}{~\ensuremath{\text{MeV}\! / c^2}}
\newcommand{\gevcc}{~\ensuremath{\text{GeV}\! / c^2}}
\newcommand{\tenpow}[2][]{%
  \ifthenelse{\equal{#1}{}}
  {\ensuremath{10^{#2}}}
  {\ensuremath{{#1} \cdot 10^{#2}}}
}

\title{Light-Meson Spectroscopy with COMPASS}

\ShortTitle{HQL 2010}

\author{\speaker{Boris Grube} for the COMPASS Collaboration \\
  Physik-Department E18 \\ Technische Universit\"at M\"unchen \\ James-Franck-Str. \\ D-85748 Garching \\ Germany \\
  E-mail: \email{bgrube@ph.tum.de}}

\abstract{%
  COMPASS is a multi-purpose fixed-target experiment at the CERN Super
  Proton Synchrotron investigating the structure and spectrum of
  hadrons. One primary goal is the search for new hadronic states, in
  particular spin-exotic mesons and glueballs. After a short pilot run
  in 2004 with a 190~GeV$/c$ $\pi^-$~beam on a \PPb~target, which
  showed a significant spin-exotic $\jpc = 1^{-+}$ resonance
  consistent with the controversial $\Ppione(1600)$, COMPASS collected
  large data samples with negative and positive hadron beams on H$_2$,
  Ni, W, and \PPb\ targets in 2008 and 2009. We present results from a
  partial-wave analysis of diffractive dissociation of 190~\gevc\
  \Ppim\ into \threepion\ final states on \PPb\ and H$_2$ targets with
  squared four-momentum transfer in the range $0.1 < t' <
  1\gevcsq$. This reaction provides clean access to the light-quark
  meson spectrum up to masses of 2.5~\gevcc. A first comparison of the
  data from \PPb\ and H$_2$ target shows a strong target dependence of
  the production strength of states with spin projections $M = 0$ and
  $1$ relative to the $\Patwo(1320)$. The 2004 \PPb\ data were also
  analyzed in the region of small squared four-momentum transfer $t' <
  \tenpow{-2}\gevcsq$, where we observe interference of diffractive
  production and photoproduction in the Coulomb-field of the \PPb\
  nucleus.  }

\FullConference{The Xth Nicola Cabibbo International Conference on Heavy Quarks and Leptons,\\
		October 11-15, 2010\\
		Frascati (Rome) Italy}

\begin{document}

\section{Introduction}
\label{sec:intro}

The COmmon Muon and Proton Apparatus for Structure and Spectroscopy
(COMPASS)~\cite{compass} is a fixed-target experiment at the CERN
Super Proton Synchrotron (SPS). It is a two-stage spectrometer that
covers a wide range of scattering angles and particle momenta with
high angular resolution. It is equipped with hadronic and
electromagnetic calorimeters so that COMPASS can reconstruct final
states with charged as well as neutral particles. The target is
surrounded by a Recoil Proton Detector (RPD) that measures the time of
flight of the recoil protons. COMPASS uses the M2 beam line of the SPS
which can deliver secondary hadron beams with a momentum of up to
300\gevc\ and a maximum intensity of
\tenpow[5]{7}~$\text{s}^{-1}$. The negative hadron beam consists of
96.0~\% \Ppim, 3.5~\% \PKm, and 0.5~\% \Ppbar. Two ChErenkov
Differential counters with Achromatic Ring focus (CEDAR) upstream of
the target are used to identify the incoming beam particles.

Its large acceptance, high resolution, and high-rate capability make
the COMPASS experiment an excellent device to study the spectrum of
light mesons in diffractive and central production up to masses of
about 2.5~GeV$/c^2$. Since COMPASS is able to measure final states
with charged as well as neutral particles, resonances can be studied
in many different reactions and decay channels.

\begin{figure}[b]
  \begin{center}
    \begin{minipage}[t]{0.475\textwidth}
      \includegraphics[width=\textwidth]{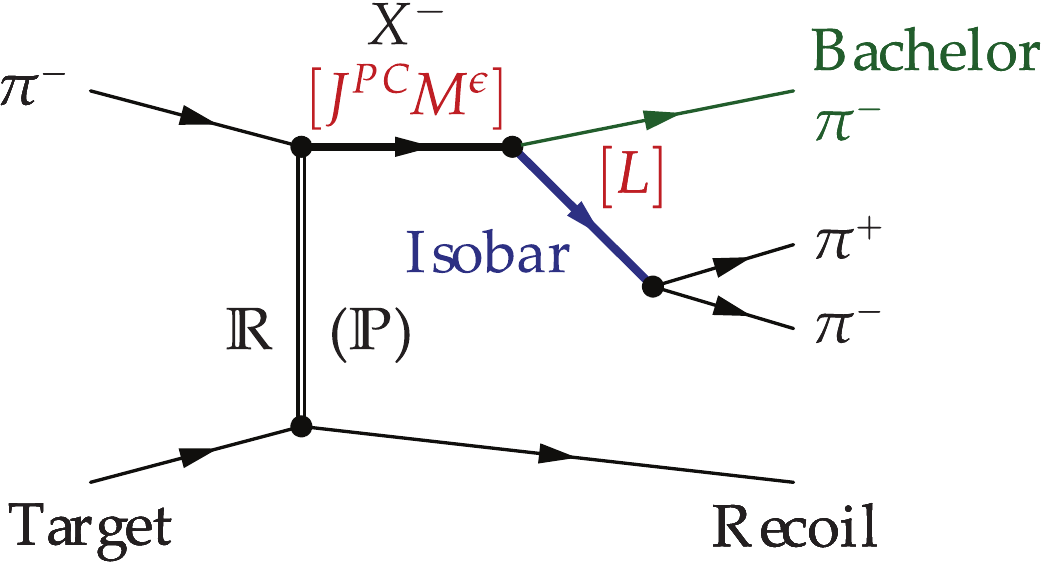}
      \caption{Diffractive production of a resonance $X^-$ via
        $t$-channel Reggeon exchange and its decay into the
        \threepion\ final state as described in the isobar model.}
      \label{fig:isobar}
    \end{minipage} \hfill
    \begin{minipage}[t]{0.475\textwidth}
      \includegraphics[width=\textwidth]{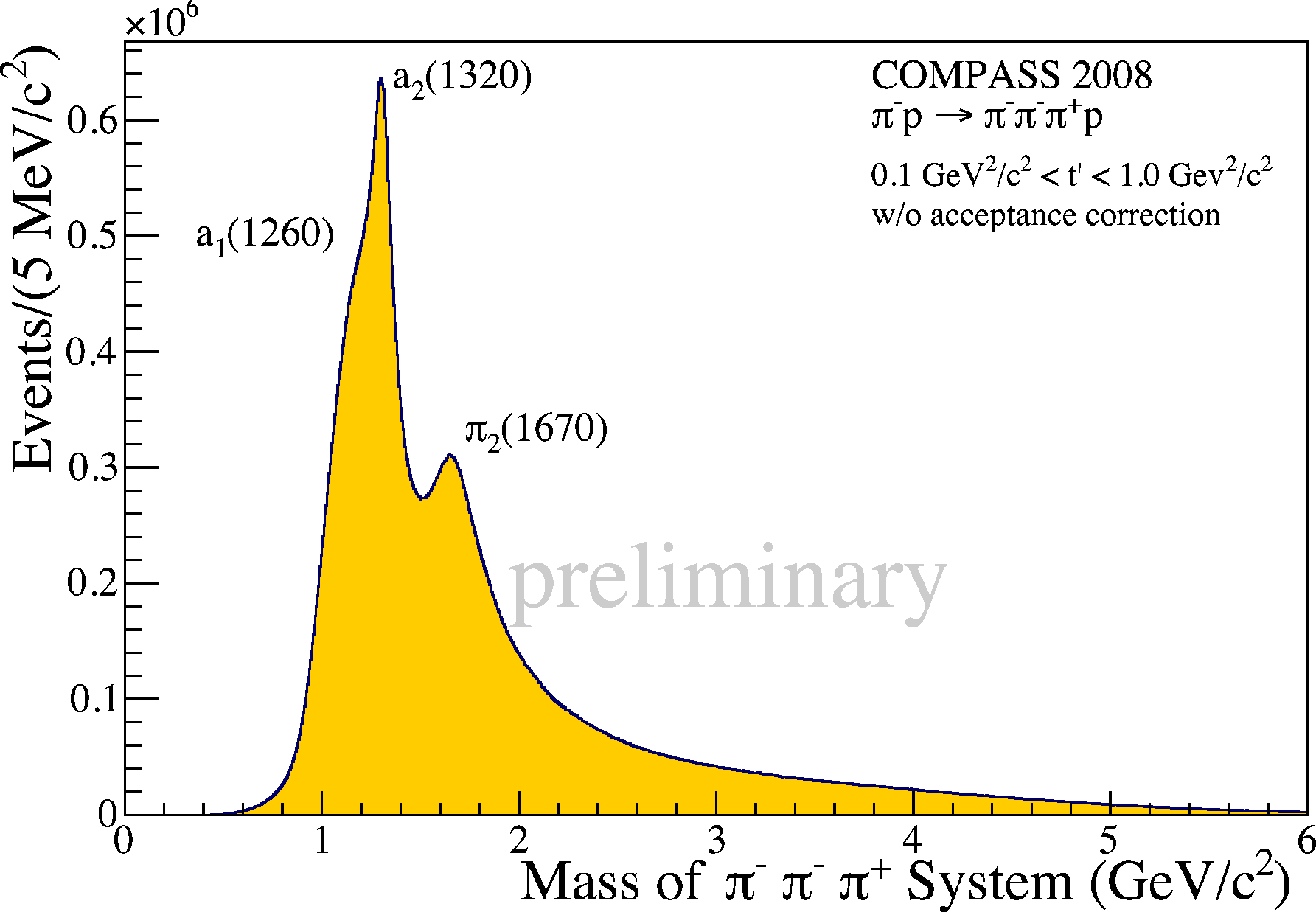}
      \caption{\threepion\ invariant mass distribution of the 2008
        data sample on H$_2$ target with $0.1 < t' < 1.0\gevcsq$.}
      \label{fig:mass}
    \end{minipage}
  \end{center}
\end{figure}

During a pilot run in 2004 and subsequent data taking periods in 2008
and 2009 COMPASS has acquired large data sets of diffractive
dissociation of 190\gevc\ \Ppim\ on H$_2$, Ni, W, and \PPb\
targets. In these events the beam pion is excited to some resonance
$X^-$ via $t$-channel Reggeon exchange with the target
(cf. \figref{isobar}). At 190\gevc\ the process is dominated by
Pomeron exchange, so that isospin and $G$-parity of the intermediate
state $X^-$ are that of the beam pion.

In 2004 the trigger selected one incoming and at least two outgoing
charged particles, whereas in 2008 a signal from the recoil proton was
required in the RPD. In the offline event selection diffractive events
were enriched by an exclusivity cut of $\pm 4$~GeV around the nominal
beam energy.

\begin{figure}[t]
  \begin{center}
    \begin{minipage}[t]{0.475\textwidth}
      \includegraphics[width=\textwidth]{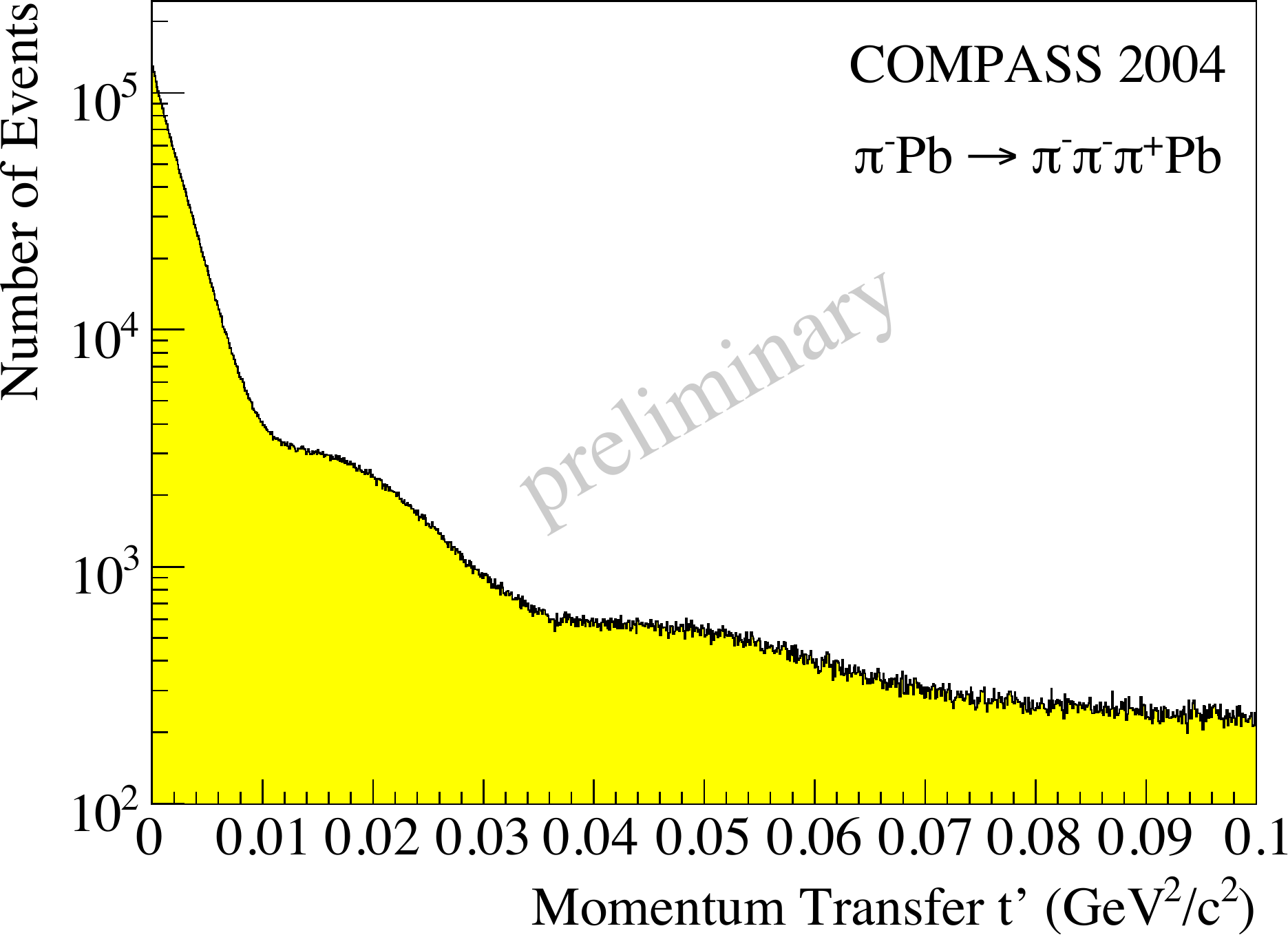}
      \caption{$t'$ spectrum of the 2004 data on a \PPb\ target: At
        very small momentum transfer of $t' < \tenpow{-3}\gevcsq$ (not
        visible in this plot) photoproduction in the Coulomb field of
        the target nucleus contributes. For $t'$ up to about
        \tenpow{-2}\gevcsq\ the spectrum can be described by an
        exponential distribution. At larger $t'$ the data exhibit a
        diffraction pattern.}
      \label{fig:tPrimePb}
    \end{minipage} \hfill
    \begin{minipage}[t]{0.475\textwidth}
      \includegraphics[width=\textwidth]{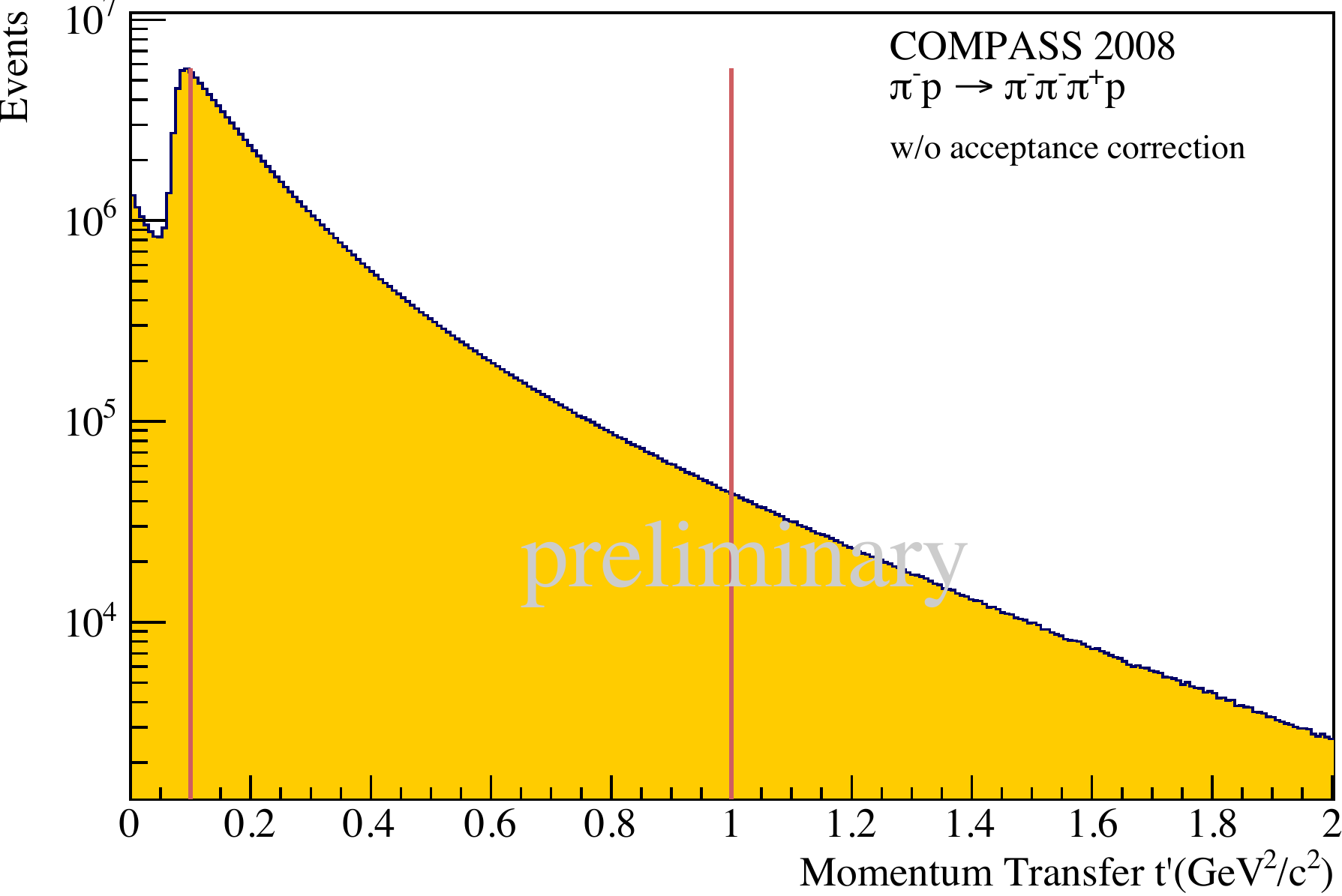}
      \caption{$t'$ distribution of the 2008 H$_2$ data: The data
        were recorded by triggering on the recoil proton so that
        events with $t'$ below about $0.1\gevcsq$ are suppressed.}
      \label{fig:tPrimeH2}
    \end{minipage}
  \end{center}
\end{figure}

Diffractive reactions are known to exhibit a rich spectrum of produced
states and are characterized by two kinematic variables: the square of
the total center-of-mass energy and the squared four-momentum transfer
from the incoming beam particle to the target, $t = (p_\text{beam} -
p_X)^2$. It is customary to use the variable $t' = \abs{t} -
\abs{t}_\text{min}$ instead of $t$, where $\abs{t}_\text{min}$ is the
minimum value of $\abs{t}$ for a certain three-pion invariant
mass. \Figrefs{tPrimePb} and \figrefshort{tPrimeH2} show the $t'$
distributions of the 2004 data sample with a \PPb\ target and that of
the 2008 data on H$_2$, respectively.

\Figref{mass} shows the \threepion\ invariant mass distribution of the
2008 data sample. It exhibits clear structures in the mass regions of
the well-known resonances $\Paone(1260)$, $\Patwo(1320)$, and
$\Ppitwo(1670)$. In order to find and disentangle the various
resonances in the data, a Partial-Wave Analysis (PWA) was
performed. In the PWA the isobar model~\cite{isobar} is used to
decompose the decay $X^- \to \threepion$ into a chain of successive
two-body decays as shown in \figref{isobar}: The $X^-$ with quantum
numbers \jpc\ and spin projection $M^\epsilon$ decays into a di-pion
resonance, the so-called isobar, and a bachelor pion. The isobar has
spin $S$ and a relative orbital angular momentum $L$ with respect to
$\Ppim_\text{bachelor}$. A partial wave is thus defined by $\jpc
M^\epsilon[\text{isobar}]L$, where $\epsilon = \pm 1$ is the
reflectivity~\cite{reflectivity}.

The production amplitudes are determined by extended maximum
likelihood fits performed in 40\mevcc\ wide bins of the three-pion
invariant mass $m_X$. In these fits no assumption is made on the
produced resonances $X^-$ other then that their production strengths
are constant within a $m_X$ bin. The PWA model includes five \twopion\
isobars~\cite{compassExotic}: $(\Ppi\Ppi)_\text{$S$-wave}$, $\Pr(770)$,
$\Pfzero(980)$, $\Pftwo(1270)$, and $\Prthree(1690)$.  They were
described using relativistic Breit-Wigner line shape functions
including Blatt-Weisskopf barrier penetration
factors~\cite{bwFactor}. For the \twopion\ $S$-wave we use the
parameterization from~\cite{vesSigma} with the $\Pfzero(980)$
subtracted from the elastic \Ppi\Ppi\ amplitude and added as a
separate Breit-Wigner resonance.

\pagebreak
\section{PWA in Large-Momentum-Transfer Region}
\label{sec:hight}

The PWA model for the squared four-momentum transfer region $0.1 < t'
< 1\gevcsq$ (cf. \figref{tPrimeH2}) consists of 41~partial waves with
$J \leq 4$ and $M \leq 1$ plus one incoherent isotropic background
wave. Because in the chosen $t'$ range the beam predominantly scatters
off the individual nucleons in the target nucleus, a rank-two
spin-density matrix is used. This accounts for spin-flip and
spin-non-flip amplitudes at the target vertex. In order to describe
the data, mostly positive reflectivity waves are needed which
corresponds to production with natural parity exchange.

The three main waves \wavespec{1}{+}{+}{0}{+}{\Pr\Ppi}{S},
\wavespec{2}{+}{+}{1}{+}{\Pr\Ppi}{D}, and
\wavespec{2}{-}{+}{0}{+}{\Pftwo\Ppi}{S} contain resonant structures
that correspond to the $\Paone(1260)$, $\Patwo(1320)$, and
$\Ppitwo(1670)$, respectively. The resonance parameters extracted from
the 2004 \PPb\ data are in good agreement with the PDG
values~\cite{compassExotic}. In addition the data exhibit a resonant
peak around 1660\mevcc\ in the spin-exotic
\wavespec{1}{-}{+}{1}{+}{\Pr\Ppi}{P} wave that is consistent with the
disputed $\Ppione(1600)$~\cite{compassExotic}.

\begin{figure}[t]
  \begin{center}
    \includegraphics[width=0.44\textwidth]{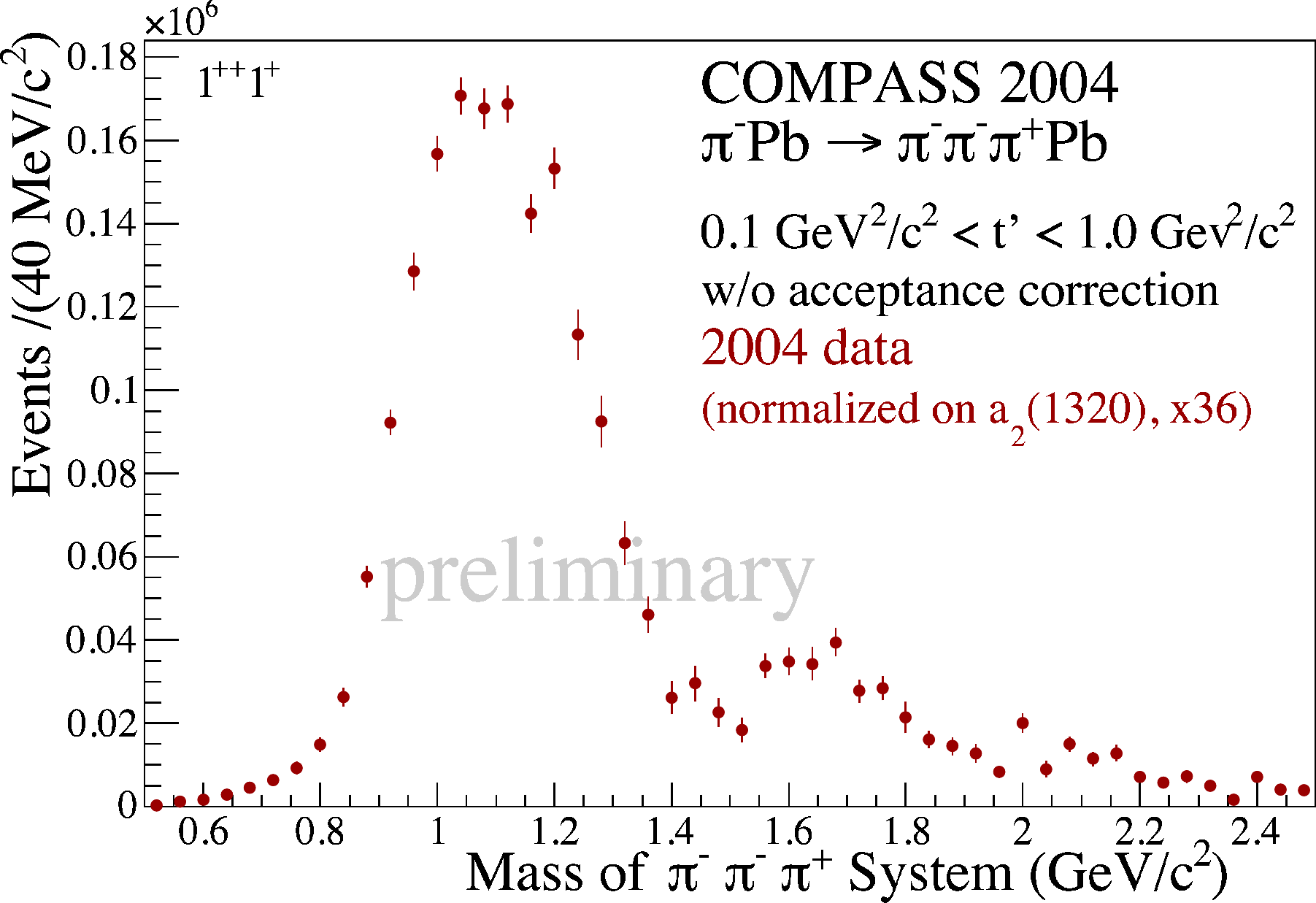}
    \includegraphics[width=0.44\textwidth]{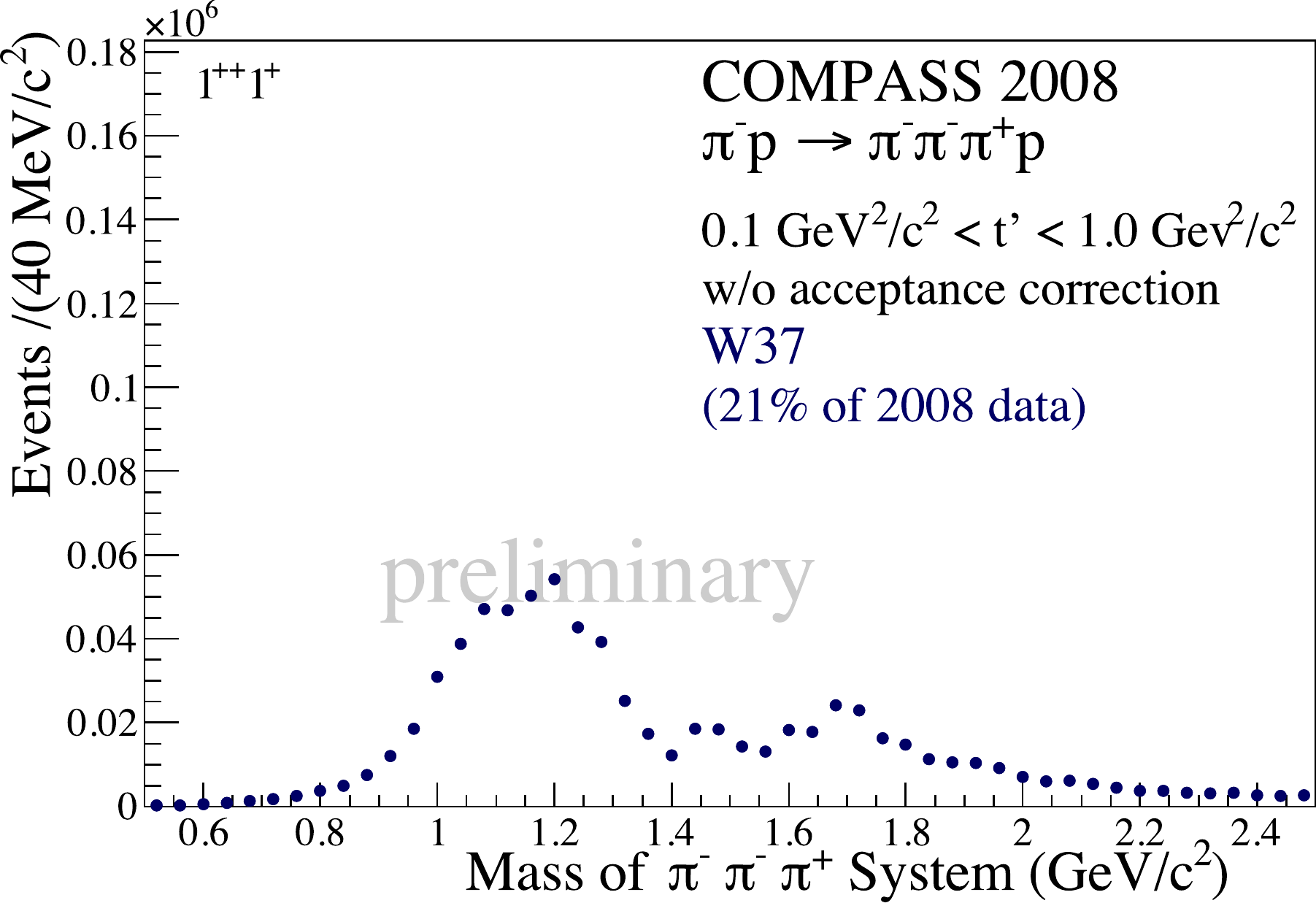} \\
    \includegraphics[width=0.44\textwidth]{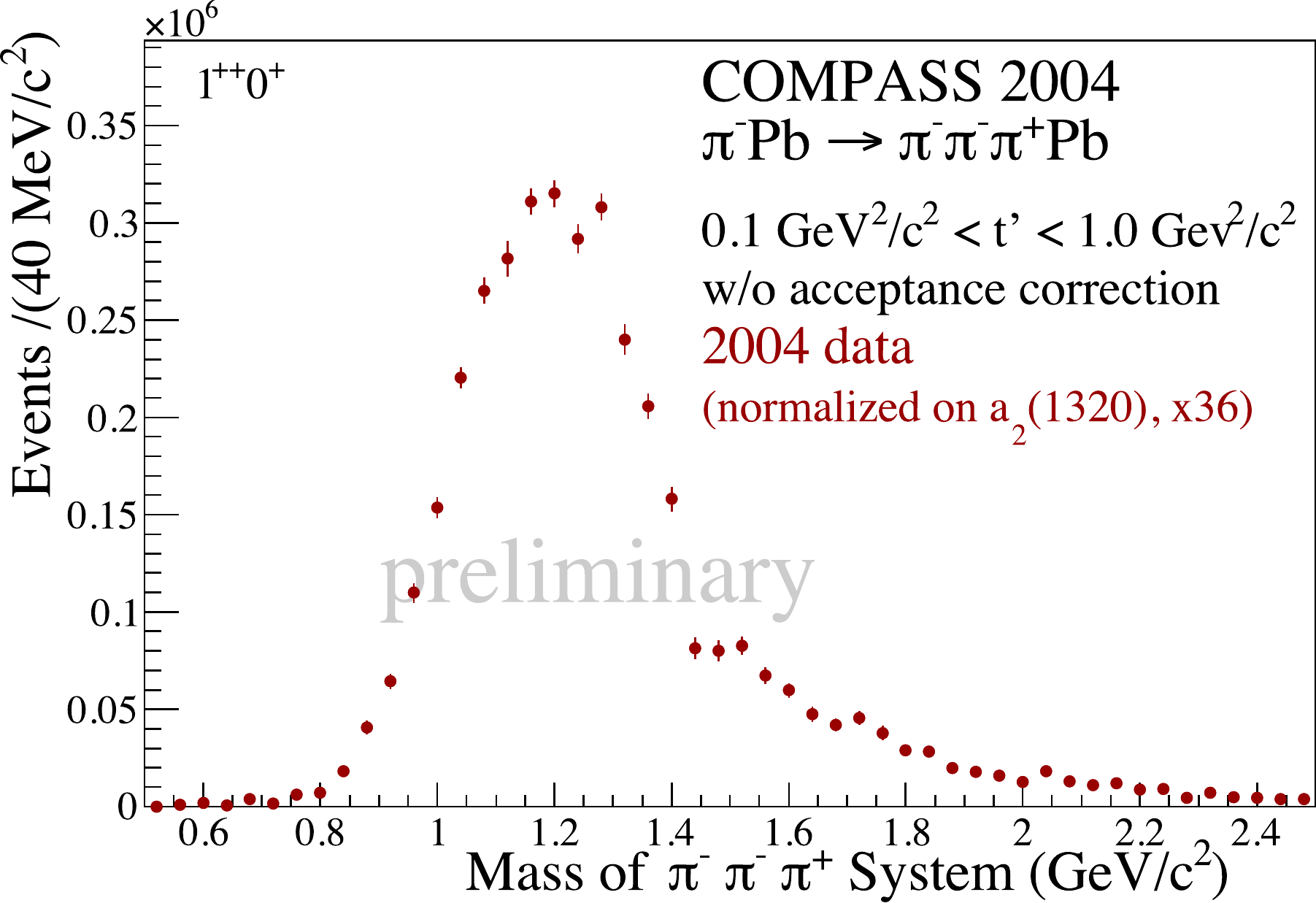}
    \includegraphics[width=0.44\textwidth]{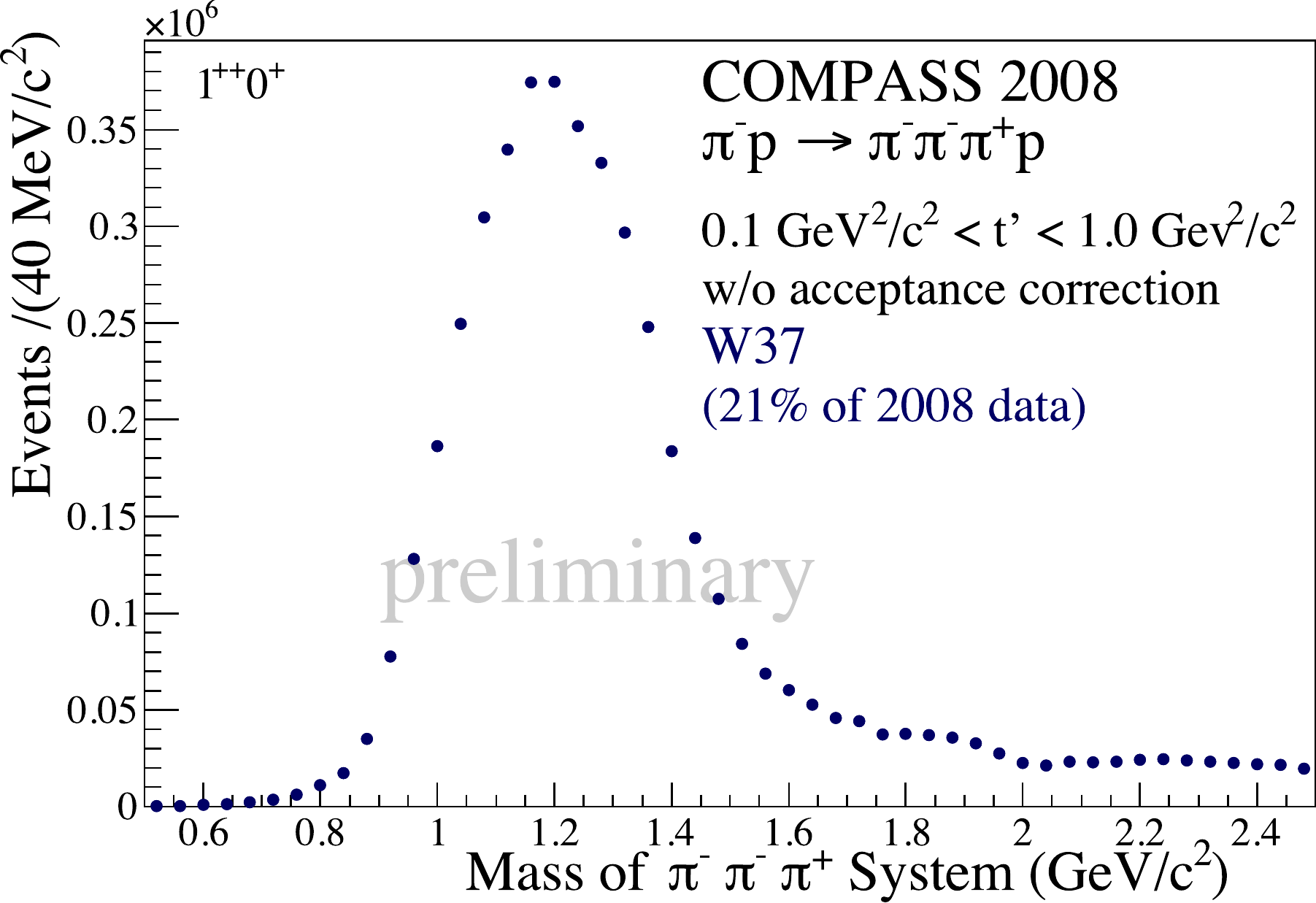} \\
    \includegraphics[width=0.44\textwidth]{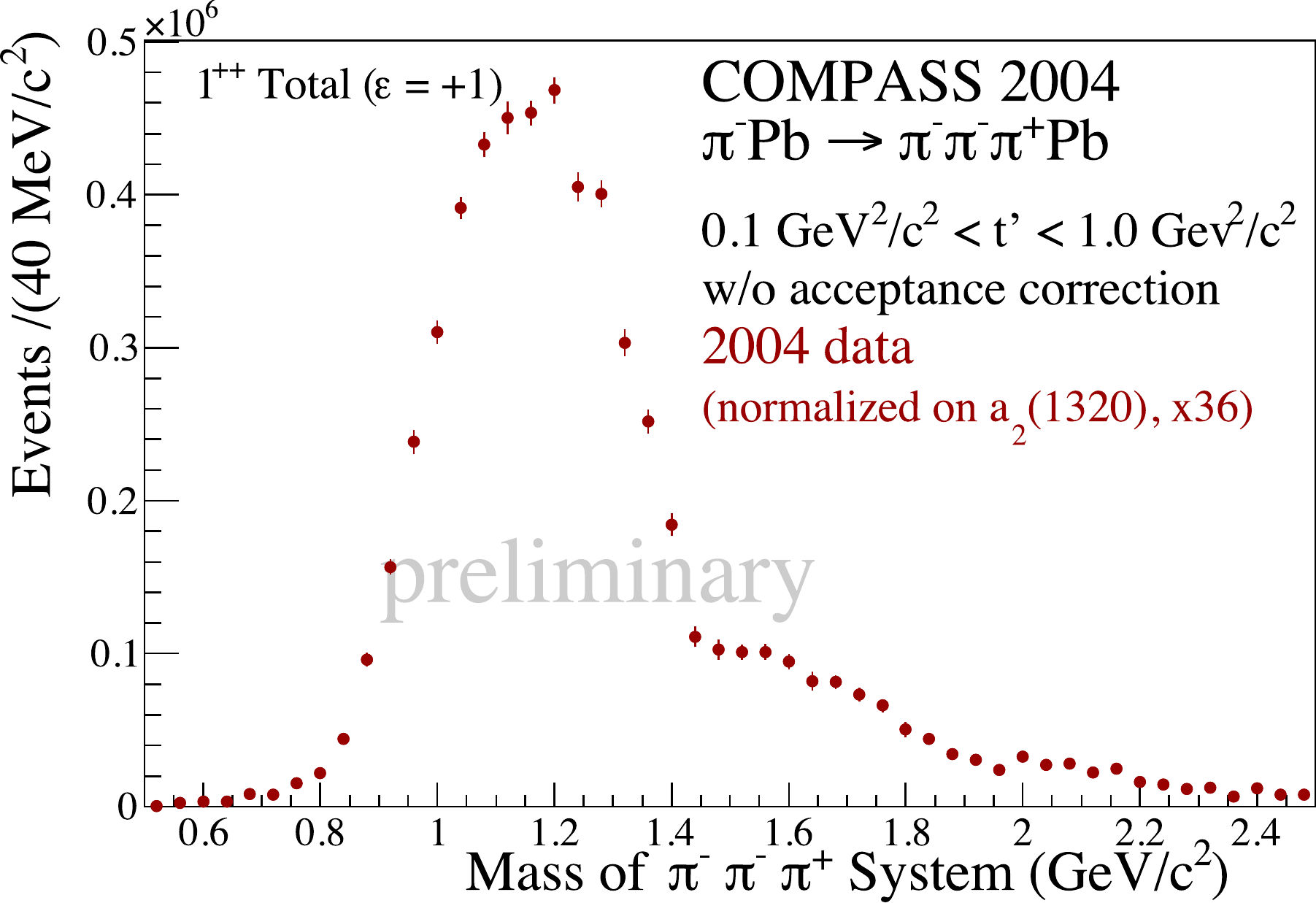}
    \includegraphics[width=0.44\textwidth]{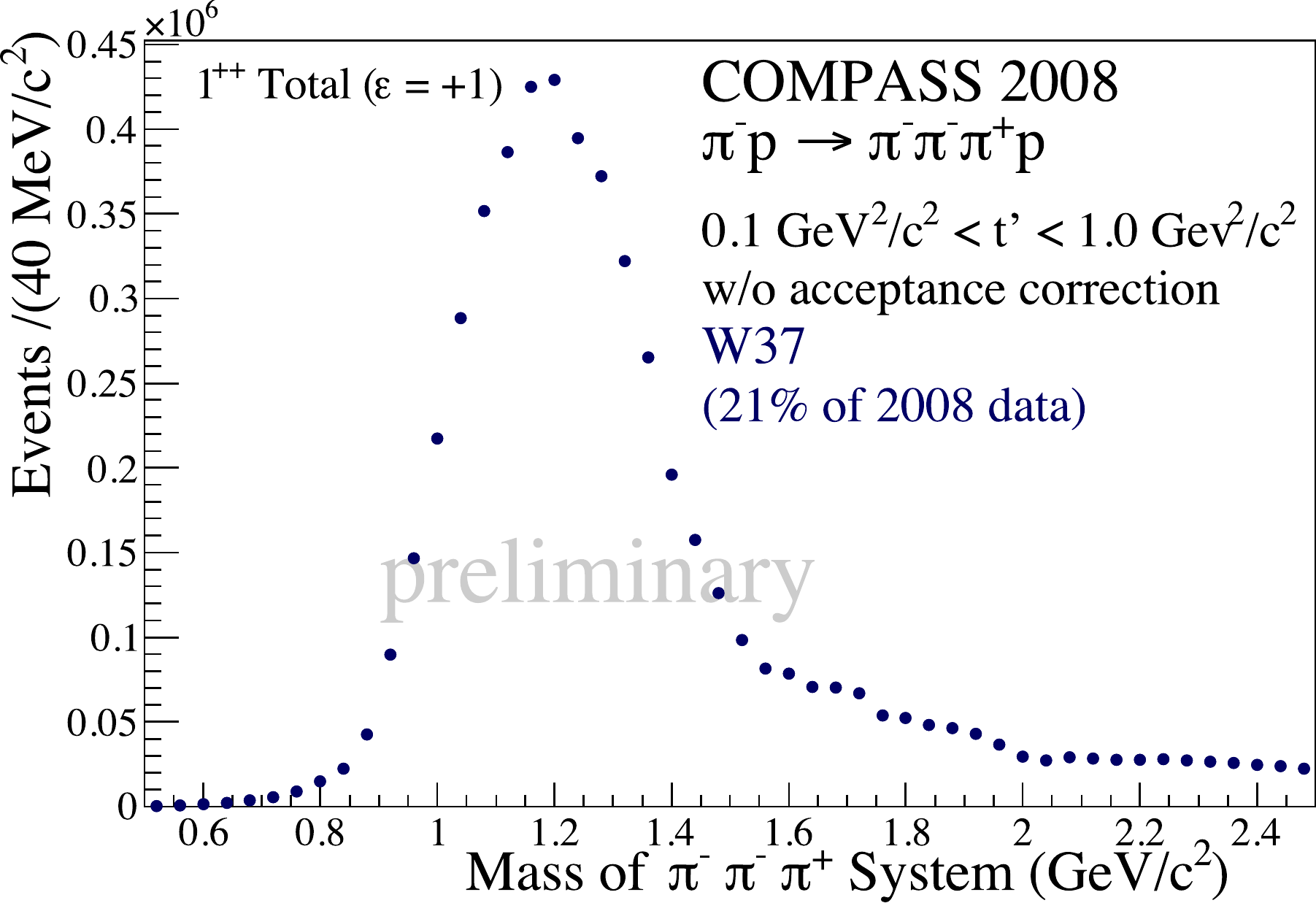} \\
    \caption{Normalized intensity sums of the $\jpc = 1^{++}$ partial
      waves for spin projection quantum numbers $M = 1$ (top) and $M =
      0$ (middle). The sum over all $M$ is shown in the bottom
      plots. The left column shows 2004 data with \PPb\ target, the
      right column 2008 data on H$_2$ target. The wave intensities are
      dominated by a broad structure around 1.2\gevcc\ which is the
      $\Paone(1260)$.}
    \label{fig:a1_MDep}
  \end{center}
\end{figure}

A first comparison of the 2008 H$_2$ data with the 2004 \PPb\ data
without acceptance corrections shows a surprisingly large dependence
on the target material. The two data sets are normalized to the narrow
$\Patwo(1320)$ resonance in the \wavespec{2}{+}{+}{1}{+}{\Pr\Ppi}{D}
wave. The H$_2$ data exhibit a strong suppression of $M = 1$ waves
relative to the $\Patwo(1320)$, whereas the corresponding $M = 0$
waves are enhanced such that the intensity sum over $M$ remains about
the same for both target materials. \Figref{a1_MDep} shows this
effect for the $\Paone(1260)$ peak in the $\jpc = 1^{++}$ waves. This
will be studied further using the data taken with Ni and W targets.

\section{PWA in Small-Momentum-Transfer Region}
\label{sec:lowt}

In the small-momentum-transfer region $t' < \tenpow{-2}\gevcsq$
(cf. \figref{tPrimePb}) two production mechanisms contribute: At
impact parameters larger than the radius of the target nucleus, which
corresponds to very low $t'$, electromagnetic interactions
dominate. In these processes the incoming beam pions scatter off the
quasi-real virtual photons that surround the heavy target
nucleus~\cite{primakoff}. For $t'$ larger than about
\tenpow{-3}\gevcsq\ the strong interaction in the form of Pomeron
exchange becomes dominant.

\begin{figure}[t]
  \begin{center}
    \includegraphics[width=0.329\textwidth]{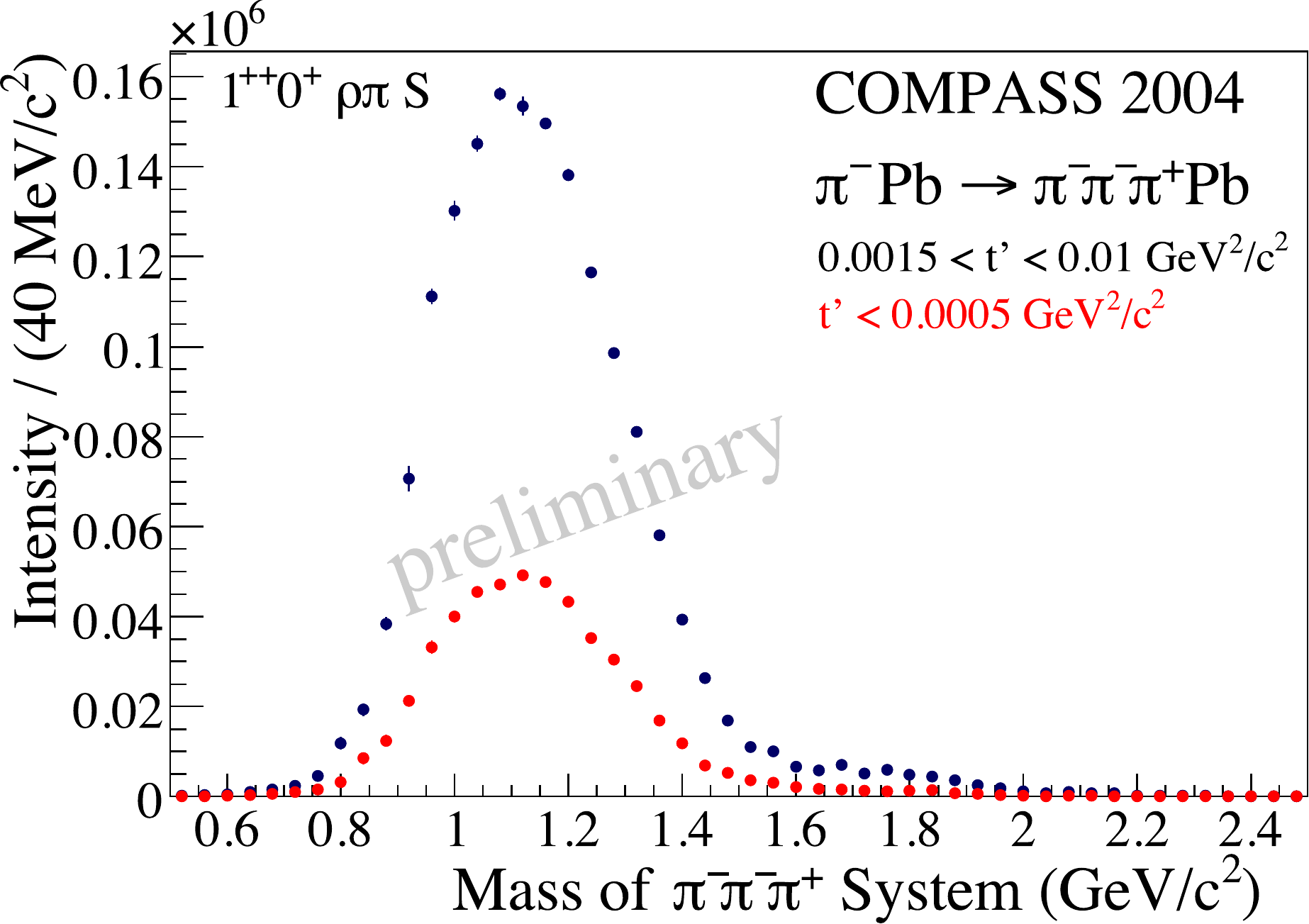}
    \includegraphics[width=0.329\textwidth]{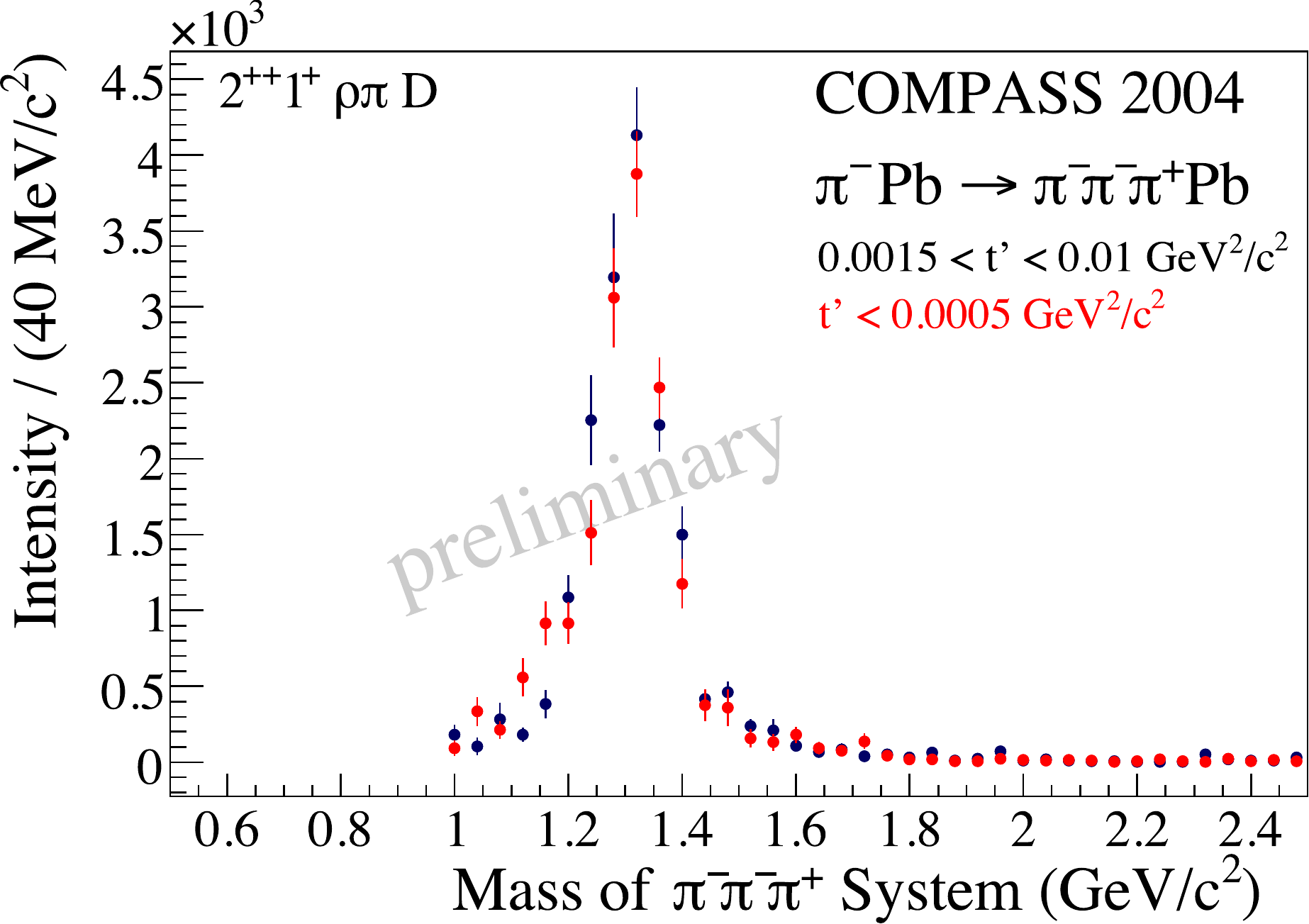}
    \includegraphics[width=0.329\textwidth]{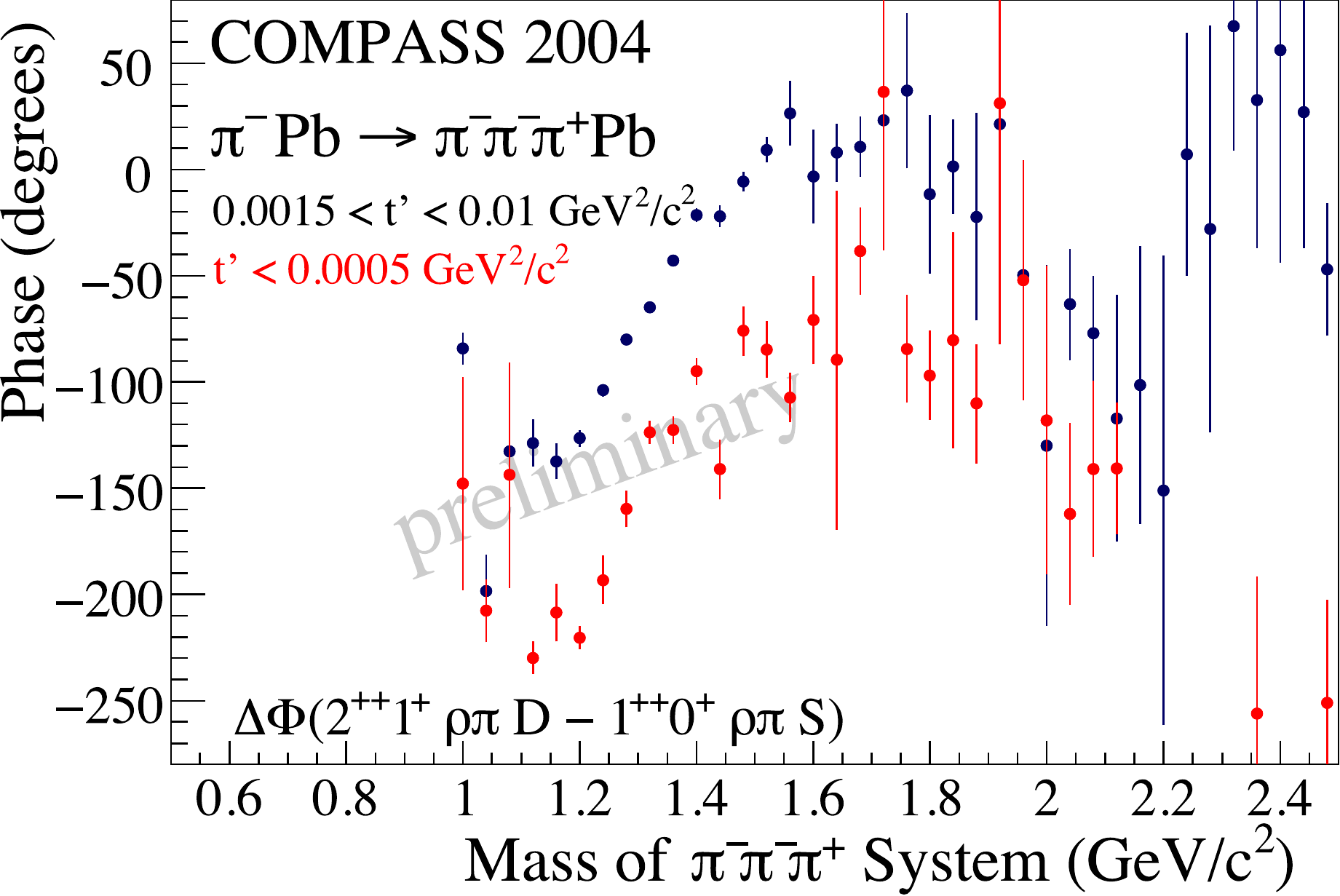}
    \caption{Intensities of the \wavespec{1}{+}{+}{0}{+}{\Pr\Ppi}{S}
      (left) and \wavespec{2}{+}{+}{1}{+}{\Pr\Ppi}{D} waves (center)
      plus their phase difference (right) as a function of the
      \threepion\ mass in two bins of the squared four-momentum
      transfer $t'$: Blue points represent the ``low-$t'$'' region
      $\tenpow[1.5]{-3} < t' < \tenpow{-2}\gevcsq$, red points the
      ``Primakoff'' region $t' < \tenpow[0.5]{-3}\gevcsq$.}
    \label{fig:a1a2_tBins}
  \end{center}
\end{figure}

The data are fitted using a PWA model that consists of 37~waves (plus
an incoherent isotropic background wave) and a rank-two spin-density
matrix. In a first analysis the $t'$ range was further subdivided into
a ``low-$t'$'' region with $\tenpow[1.5]{-3} < t' <
\tenpow{-2}\gevcsq$ and a ``Primakoff'' region with $t' <
\tenpow[0.5]{-3}\gevcsq$. The low-$t'$ region is dominated by
diffractive production processes, whereas in the Primakoff region also
photoproduction contributes. This effect is enhanced for waves with
spin projection $M = 1$, because the diffractive production strength
is proportional to $(t')^M\, \exp(-b t')$. \Figref{a1a2_tBins} shows
the behavior of the two main waves: the
\wavespec{1}{+}{+}{0}{+}{\Pr\Ppi}{S} wave, which mainly consists of
the $\Paone(1260)$ resonance, has $M = 0$ so that diffractive
production is still dominant in the Primakoff region. The
$\Patwo(1320)$ resonance in the \wavespec{2}{+}{+}{1}{+}{\Pr\Ppi}{D}
wave, however, has $M = 1$ so that in the Primakoff region diffractive
production of this resonance is suppressed whereas photoproduction is
enhanced. As \figref{a1a2_tBins} (center) shows, the $t'$ bins were
chosen such that the number of $\Patwo(1320)$ in the Primakoff region
is similar to that in the low-$t'$ region. The phase motion of the
narrow $\Patwo(1320)$ in the large-mass tail of the $\Paone(1260)$ is
similar in both $t'$ regions except for an offset of about
90\textdegree\ (cf. \figref{a1a2_tBins} right). This is due to the
different production phases of the $\Patwo(1320)$ introduced by photon
and Pomeron exchange, respectively.

\begin{figure}[t]
  \begin{center}
    \includegraphics[width=0.44\textwidth]{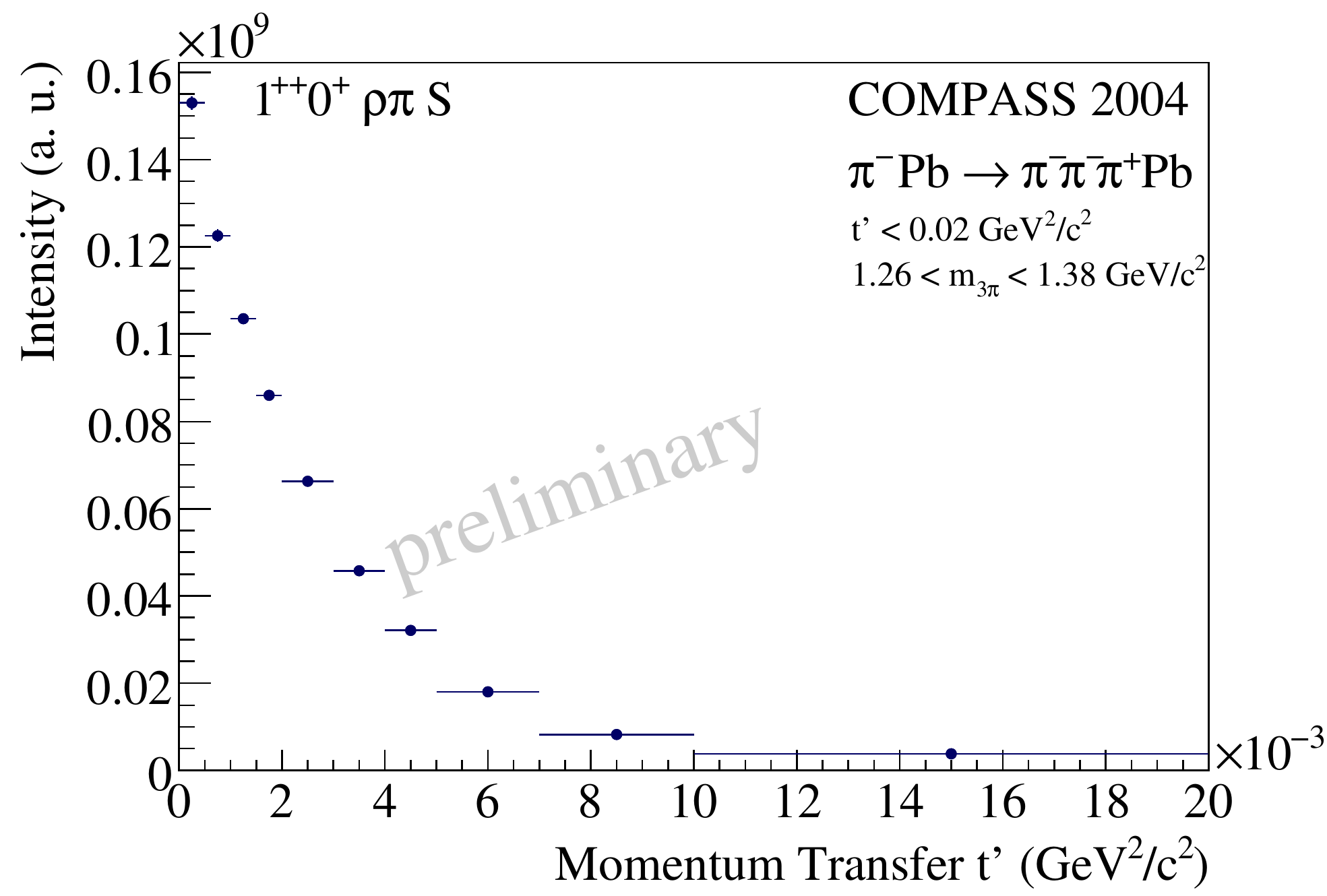}
    \includegraphics[width=0.44\textwidth]{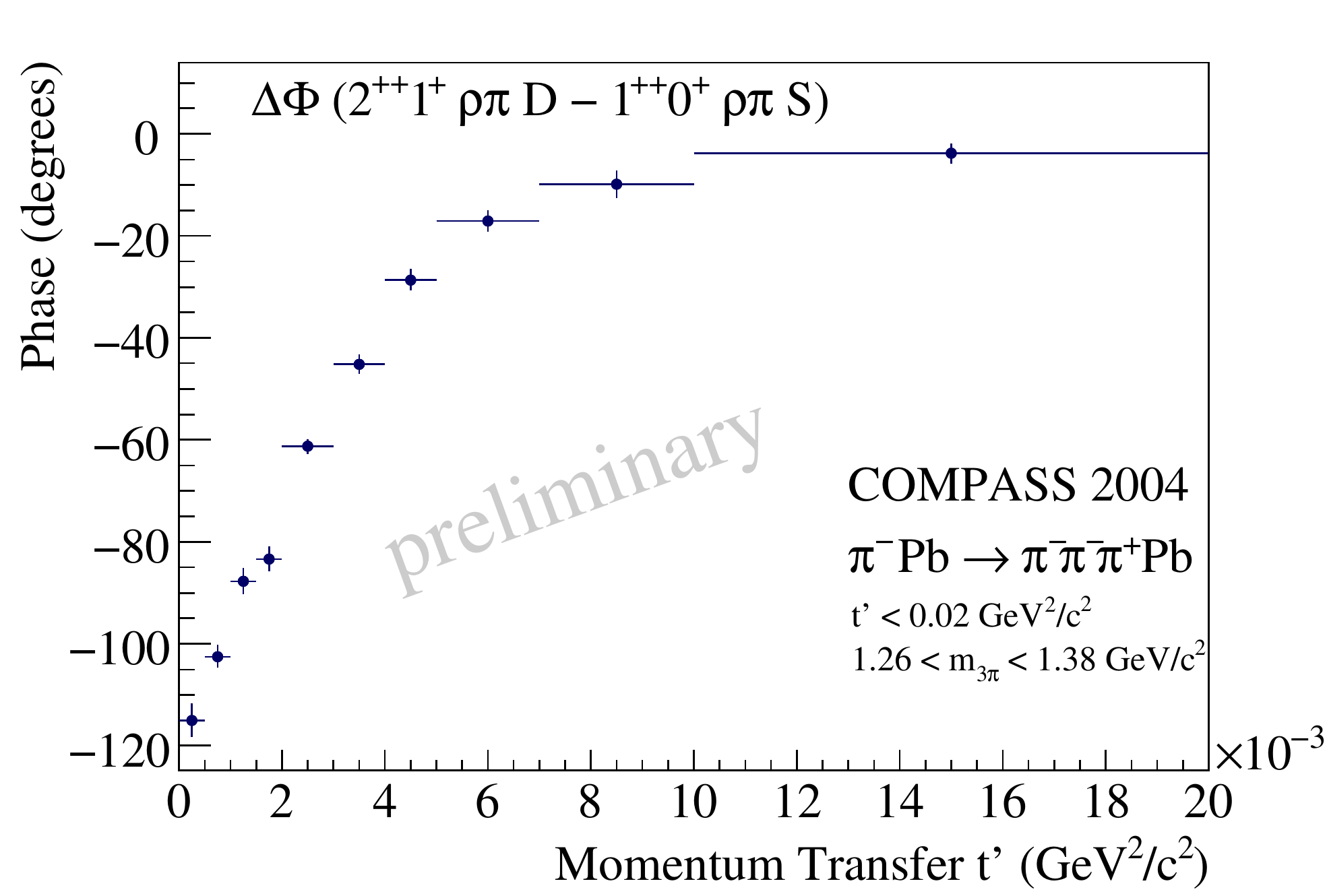} \\
    \hspace*{-0.9em}\includegraphics[width=0.44\textwidth]{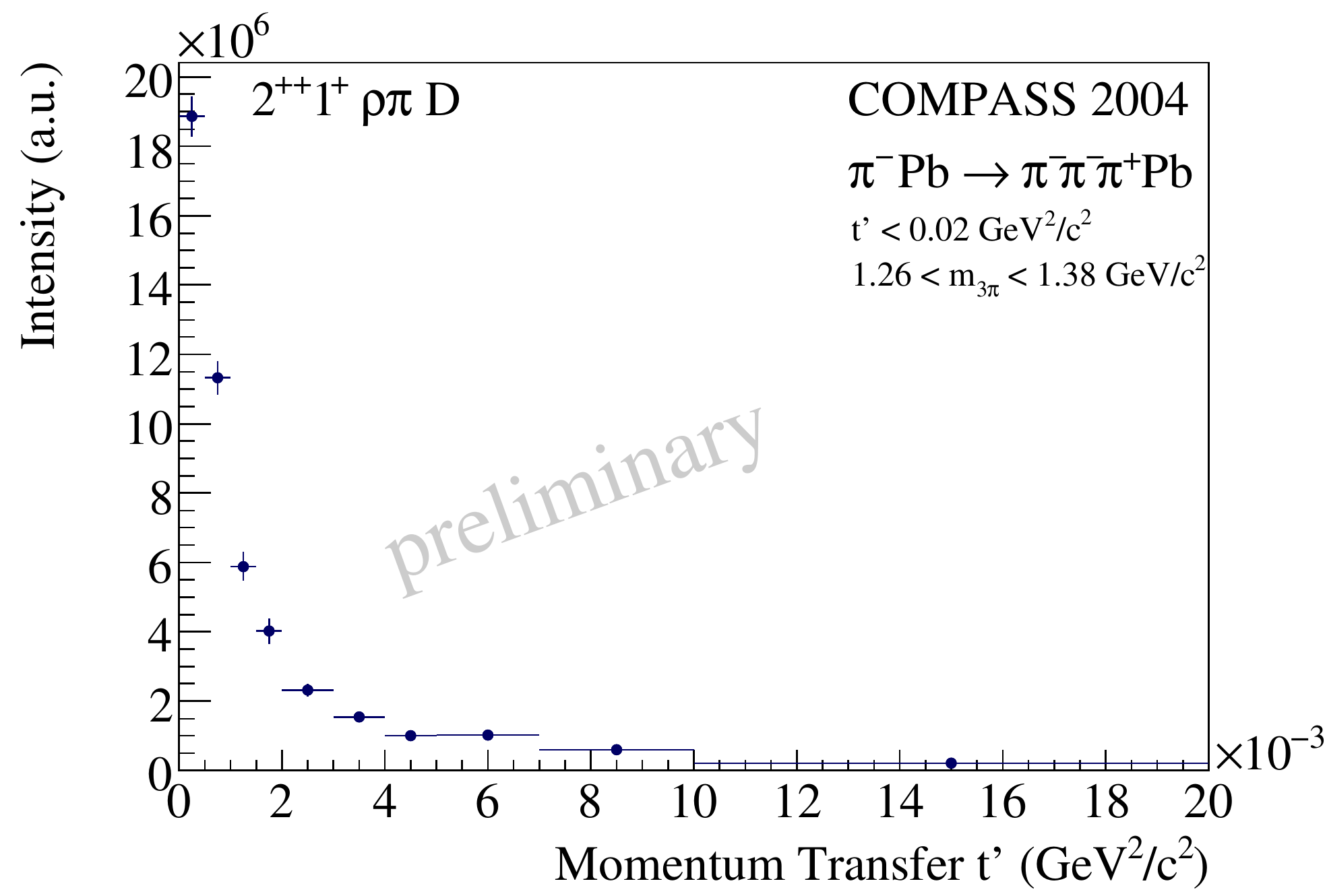}
    \hspace*{0.65em}\includegraphics[width=0.4\textwidth]{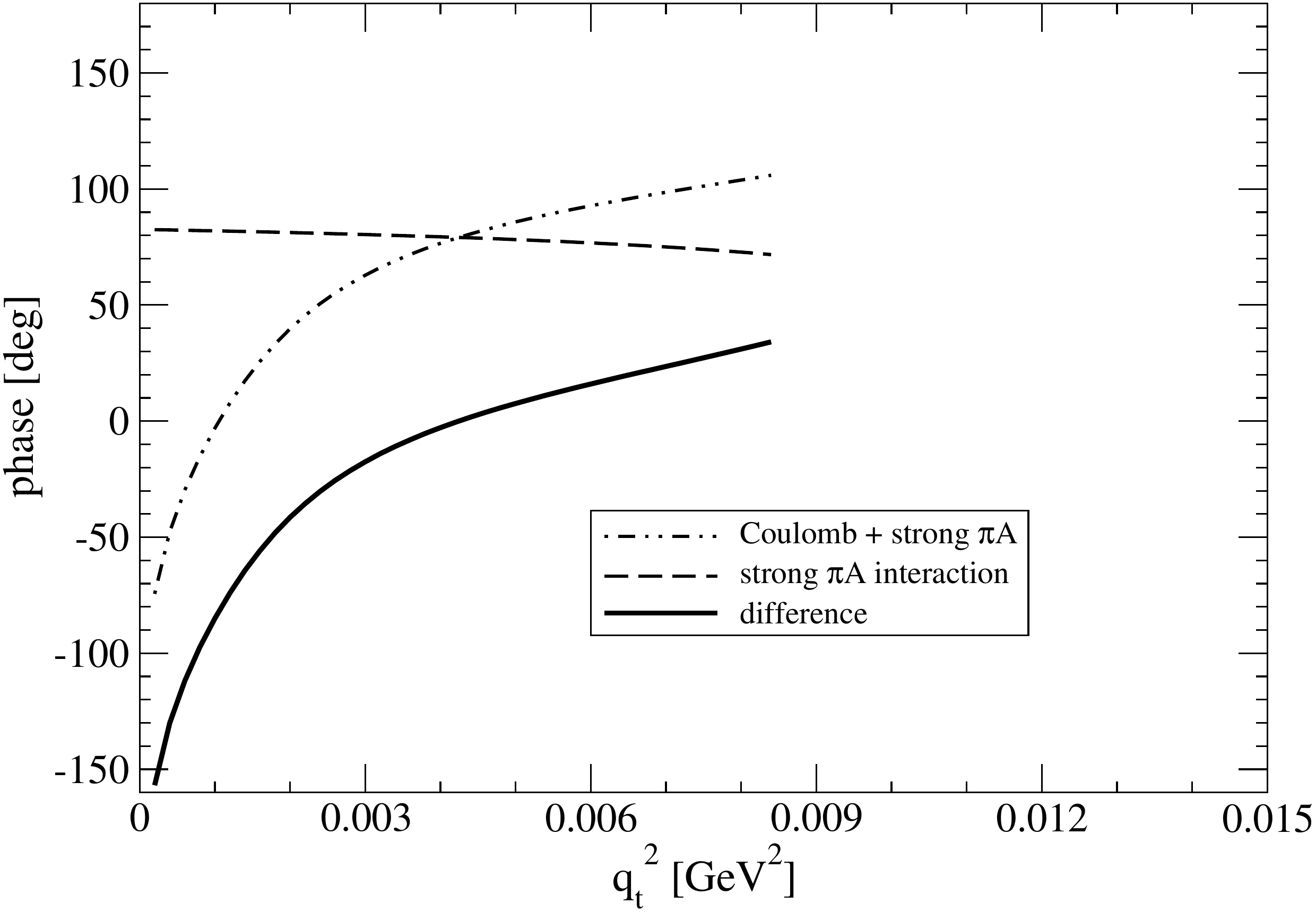}
    \caption{Intensities of the \wavespec{1}{+}{+}{0}{+}{\Pr\Ppi}{S}
      (top-left) and \wavespec{2}{+}{+}{1}{+}{\Pr\Ppi}{D} waves
      (bottom-left) plus their relative phase (top-right) as a
      function of the squared four-momentum transfer $t'$ in a mass
      range around the $\Patwo(1320)$ mass. The phase difference
      between the strong \Ppi\PA\ interaction and the Coulomb
      interaction as predicted by a Glauber model calculation is shown
      in the lower right plot.}
    \label{fig:a1a2_tDep}
  \end{center}
\end{figure}

In order to study the transition between the two $\Patwo(1320)$
production mechanisms, a PWA in bins of $t'$ was performed in a wide
mass region around the $\Patwo(1320)$ (see \figref{a1a2_tDep}). The
resulting $t'$ spectrum of the $\Paone(1260)$ has a slope compatible
with diffractive production (cf. \figref{tPrimePb}), whereas the
$\Patwo(1320)$ spectrum features a steep rise at low $t'$ as expected
for photoproduction, given the finite experimental resolution. The
phase difference of the $\Patwo(1320)$ and the $\Paone(1260)$ vanishes
at larger $t'$ indicating that both resonances are produced
diffractively. The $t'$ dependence of the relative phase is roughly
described by a Glauber model analogous to~\cite{glauberModel}. Further
analyses will focus on the study of the reaction $\Ppi \PPh \to 3
\Ppi$ close to threshold in order to test recent $\chi$PT
calculations~\cite{chiPT}.

\acknowledgments

This work is supported by the German
Bundesministerium f\"ur Bildung und Forschung
BMBF, the Maier-Leibnitz-Labor der LMU und TU M\"unchen, the
DFG Cluster of Excellence \emph{Origin and Structure of the Universe},
and CERN-RFBR grant 08-02-91009.

\end{document}